\documentclass{revtex4}
\usepackage{amsmath,amssymb,amsfonts}
\usepackage{algorithm}
\usepackage{tikz,pgf}
\usepackage{multirow}
\usepackage{epstopdf}
\usepackage{url,verbatim}

\newtheorem{definition}{Definition}
\begin{document}
\newcommand{\ket}[1]{\ensuremath{\left|#1\right\rangle}}
\newcommand{\bra}[1]{\ensuremath{\left\langle#1\right|}}
\newcommand\floor[1]{\lfloor#1\rfloor}
\newcommand\ceil[1]{\lceil#1\rceil}

\title{Measurement Device Independent Quantum Dialogue}
\author{Arpita Maitra\\
Centre for Theoretical Studies, \\
Indian Institute of Technology Kharagpur, \\
Kharagpur 721302, West Bengal, India.\\
	\email{arpita76b@gmail.com}}
\date{}

\begin{abstract}
Very recently, the experimental demonstration of Quantum Secure Direct Communication (QSDC) with state-of-the-art atomic quantum memory has been reported (Phys. Rev. Lett., 2017). Quantum Dialogue (QD) falls under QSDC where the secrete messages are communicated simultaneously between two legitimate parties. The successful experimental demonstration of QSDC opens up the possibilities for practical implementation of QD protocols. Thus, it is necessary to analyze the practical security issues of QD protocols for future implementation.  Since the very first proposal for QD by Nguyen (Phys. Lett. A, 2004) a large number of variants and extensions have been presented till date. However, all of those leak half of the secret bits to the adversary through classical communications of the measurement results.
 In this direction, motivated by the idea of Lo et al. (Phys. Rev. Lett., 2012), we propose a 
Measurement Device Independent Quantum Dialogue (MDI-QD) scheme which is resistant to such information leakage as well as side channel attacks. In the proposed protocol, Alice and Bob, two legitimate parties, are allowed to prepare the states only. The states are measured by an untrusted third party (UTP) who may himself behave as an adversary. We show that our protocol is secure under this adversarial model. The current protocol does not require any quantum memory and thus it is inherently robust against memory attacks. Such robustness might not be guaranteed in the QSDC protocol with quantum memory (Phys. Rev. Lett., 2017).
\end{abstract}
\maketitle

\section{Introduction}
\label{intro}
Since the pioneering result of Shor~\cite{Shor} in 1994, an immense attention had been directed towards quantum cryptography. The subdomain, quantum key distribution, achieves enormous proliferation. However, the other domains such as quantum secure direct communication (QSDC), quantum private query (QPQ) etc. are growing gradually.  

The ancestor of QSDC is quantum super dense coding~\cite{Superden}. In super dense coding two classical bits of information are communicated exploiting only one qubit. The original protocol exploited entangled states to perform this goal.

The first generation QSDC protocols~\cite{qd0,qd1,qd2,qd3,qd4,qd5,qd6,qd7,qd8} considered one direction, i.e., the classical bits were communicated from Alice to Bob. The subsequent work of~\cite{qd9} proposed bidirectional QSDC where secret messages can flow in both the directions, i.e., from Alice to Bob as well as from Bob to Alice. Such two way communication is known as Quantum Dialogue (QD). 

In~\cite{qd10}, the authors showed that the QD protocol was susceptible to intercept and resend attack and proposed a modified version of it. Since then several variants~\cite{qd11,qd12,banerjee,shukla,pathak} have been proposed till date.

Very recently, the experimental demonstration of QSDC with state-of-the-art atomic quantum memory has been reported~\cite{zhang}. As QD protocols are motivated towards providing algorithms for bi-directional QSDC, this experimental success opens up the possibility for practical QD. It is thus necessary to examine the vulnerabilities of the existing QD protocols. Unfortunately, all the existing QD protocols inherently leak some information to the adversary through the public announcement of the measurement results~\cite{qd13,qd14} (detailed calculations are shown in section~\ref{com}).

In this direction, motivated by the idea of Lo et al.~\cite{Lo}, we, here, propose a measurement device independent quantum dialogue (MDI-QD) protocol which is robust against not only this type of information leakage but also all side channel attacks. Moreover, our proposed protocol does not require any quantum memory. 

In QSDC Alice and Bob share $N$ number of entangled pairs which are initially prepared in $\ket{\phi^{+}}=\frac{1}{\sqrt{2}}(\ket{00}+\ket{11})$ state. Alice retains one half of each entangled pair and transmits other half to Bob. Alice and Bob randomly choose some entangled pairs to estimate the error in the channel. If the error lies below the pre-defined threshold, Alice encodes her qubits with messages i.e., performs the operation $I$, $\sigma_z$, $\sigma_x$ or $i\sigma_y$ on her qubits. Alice then sends her encoded qubits to Bob who performs Bell-state measurement to decode the message coming from Alice. If he obtains $\ket{\phi^{+}}$ (Alice performed $I$), he concludes that Alice's message was $00$. If it is $\ket{\phi^{-}}$ (Alice performed $\sigma_z$), he understands that it was $01$. If he gets $\ket{\psi^{-}}$ (Alice performed $\sigma_x$), the message was $10$. If he measures $\ket{\psi^{-}}$ (Alice performed $i \sigma_y$), the message was $11$.

In QSDC, due to the transmission of $N$ qubits and the above encoding operations, the entangled pairs shared between Alice and Bob are required to be stored for some time. Hence, quantum memory is necessary to accomplish the communication task in QSDC. However, in the present protocol, we assume that as soon as Eve (the UTP) gets the qubits from Alice and Bob, she measures those in Bell basis and announces the result publicly. Depending on the announcement Alice and Bob perfectly decode the information coming from their respective counter-parts. Hence, quantum memory is not an essential part here. One may bypass quantum memory to make the protocol inherently robust against any type of memory attacks.

In existing quantum dialogue protocols~\cite{qd9,qd10,qd11,qd12,banerjee,shukla,pathak}, Alice or Bob measures the states coming from their respective counter-parts at her or his end. Contrary to this, in the current proposal, we propose a scheme where the measurement devices are with Eve only, who might work as an adversary. Her motivation is to steal the information regarding the communication between two legitimate parties, namely Alice and Bob. In this initiative, we prove that our scheme is secure.

 The security of the proposed protocol depends on the following three assumptions. Firstly, we assume the inherent physical correctness of 
Quantum Mechanics. Second, we have an assumption that no information leakage takes place from the legitimate parties' (Alice and Bob) laboratories. Finally, we consider that the legitimate parties have a sufficiently good knowledge about their sources.

\section{Proposed Protocol}
In this section we present the protocol. To describe the algorithm, we need to exploit the Bell states. These are
two-qubits entangled states that can form orthogonal basis. The four 
Bell states can be written as 
$|\Phi^\pm\rangle = \frac{1}{\sqrt{2}}[|00\rangle \pm |11\rangle], \
|\Psi^\pm\rangle = \frac{1}{\sqrt{2}}[|01\rangle \pm |10\rangle].$

Our current protocol is divided into two parts. We compose two different protocols (BB84~\cite{bb84} and a modified version of MDI-QKD~\cite{Lo}) to propose our measurement device independent quantum dialogue protocol. One may replace BB84 part by MDI-QKD also. We have a brief discussion regarding this issue in section~\ref{con}. 

In the first part, Alice and Bob perform BB84~\cite{bb84} protocol to establish a secret key of $m$ bits between themselves. This key will be used to prepare the states 
at both ends. We denote this key by $b$. When $b_j = 0$, $j\in \{1,\cdots m\}$, Alice (Bob) prepares her (his) state in $\{\ket{0},\ket{1}\}$ basis. When $b_j=1$, Alice (Bob) prepares her (his) state in $\{\ket{+},\ket{-}\}$ basis. Here, $\ket{+}=\frac{1}{\sqrt{2}}(\ket{0}+\ket{1})$ and $\ket{-}=\frac{1}{\sqrt{2}}(\ket{0}-\ket{1})$. Now, Alice (Bob) chooses a random stream $a$ ($a'$) of $m$ bits. The encoding algorithm is presented in Algorithm $\bf{1}$.

\begin{algorithm}[htbp]
\label{algo_1}
{\centering
\begin{enumerate}
\item When $a_j (a'_j)=0$ and $b_j=0$, Alice (Bob) prepares $\ket{0}$.
\item When $a_j (a'_j)=1$ and $b_j=0$, Alice (Bob) prepares $\ket{1}$.
\item When $a_j (a'_j)=0$ and $b_j=1$, Alice (Bob) prepares $\ket{+}$.
\item When $a_j (a'_j)=1$ and $b_j=1$, Alice (Bob) prepares $\ket{-}$.
 \end{enumerate}
 }
 \caption{Algoritm for encoding}
\end{algorithm}

Alice (Bob) sends her (his) qubit to the UTP (Eve). UTP measures the two particles (one from Alice and one from Bob) in Bell basis. Depending on the announcement by Eve 
(the result of the Bell measurement), Alice and Bob decode their information. 


From Table~\ref{vvtable}, it is clear that when Alice sends $\ket{0}$ ($\ket{1}$) and the measurement result is either $\ket{\phi^{+}}$ or $\ket{\phi^{-}}$, she concludes with probability $1$ that the state sent by Bob is $\ket{0}$ (\ket{1}). That means she comes to know with certainty that Bob wants to communicate the classical bit $0$ ($1$). When Alice prepares $\ket{0}$ (\ket{1}) and the measurement result is either $\ket{\psi^{+}}$ or $\ket{\psi^{-}}$, Alice concludes that Bob wants to communicate $1$ ($0$). If Alice prepares $\ket{+}$ ($\ket{-}$) and the measurement result is either $\ket{\phi^{+}}$ or $\ket{\psi^{+}}$, Alice concludes that Bob likes to communicate $0$ ($1$). Similarly, when Alice prepares $\ket{+}$ ($\ket{-}$) and the measurement result is $\ket{\phi^{-}}$ or $\ket{\psi^{-}}$, Bob's communicated bit must be $1$ ($0$). Similar things happen for Bob too. Hence, with probability $1$, Alice and Bob can exchange two classical bits of information at a time between themselves. 
{\scriptsize
\begin{table}[htbp]
\begin{center}
\begin{tabular}{|c|c|c|c|c|c|c|c|}
\hline
\multicolumn{2}{|c|}{Qubits sent by} & \multicolumn{4}{c|}{Probability (Eve's end)} & \multicolumn{2}{c|}{Communicated Bits}\\
\hline
Alice & Bob & $|\Phi^+\rangle$ & $|\Phi^-\rangle$ & $|\Psi^+\rangle$ & $|\Psi^-\rangle$ & by Alice & by Bob\ \\
\hline
\hline
$|0\rangle$ & $|0\rangle$ & $\frac{1}{2}$ & $\frac{1}{2}$ & 0 & 0 & 0 & 0\\
$|0\rangle$ & $|1\rangle$ & 0 & 0 & $\frac{1}{2}$ & $\frac{1}{2}$ & 0 & 1\\
$|1\rangle$ & $|0\rangle$ & 0 & 0 & $\frac{1}{2}$ & $\frac{1}{2}$ & 1 & 0\\
$|1\rangle$ & $|1\rangle$ & $\frac{1}{2}$ & $\frac{1}{2}$ & 0 & 0 & 1 & 1\\
\hline
$|+\rangle$ & $|+\rangle$ & $\frac{1}{2}$ & 0 & $\frac{1}{2}$ & 0 & 0 & 0\\
$|+\rangle$ & $|-\rangle$ & 0 & $\frac{1}{2}$ & 0 & $\frac{1}{2}$ & 0 & 1\\
$|-\rangle$ & $|+\rangle$ & 0 & $\frac{1}{2}$ & 0 & $\frac{1}{2}$ & 1 & 0\\
$|-\rangle$ & $|-\rangle$ & $\frac{1}{2}$ & 0 & $\frac{1}{2}$ & 0 & 1 & 1\\
\hline
\end{tabular}
\end{center}
\caption{Different cases in MDI QSDC} 
\label{vvtable}
\end{table}}

One can notice that in case of $\ket{\phi^{+}}$, Eve knows that the communicated bits between Alice and Bob are either $00$ or $11$. Similarly, in case of $\ket{\psi^{-}}$ those must be either $01$ or $10$.  That means in these two cases, the measurement outcomes leak $1$ bit of information to Eve. To avoid this situation, Alice and Bob will continue the QD protocol only for the cases where the measurement outcomes are $\ket{\phi^{-}}$ or $\ket{\psi^{+}}$. 

Similar to the first phase (BB84 phase) of the protocol, in this phase also, we need to check the error introduced due to eavesdropping. This is because, the adversary certainly knows which phase is being implemented
at a given time and according to that she could adapt her strategy. To resist this, Alice and Bob choose a subset of $\frac{\gamma m}{2}$ number of runs among remaining $\frac{m}{2}$ number of runs, where $\gamma$ is a small fraction. For this $\frac{\gamma m}{2}$ runs, Alice and Bob reveal their respective guesses about the communicated bits by their counter-parts and check the noise introduced in the channel. If the noise lies below the predefined threshold value, Alice and Bob will continue the protocol. Otherwise, they abort. 

To reduce the misuse of bits one could exploit the discarded bits for error estimation. However, if it opens up any security loop-hole is out of the scope here. 

One may find that after sharing the bit stream $b$, Alice and Bob can communicate $\frac{m}{2}$ bits to each other directly, i.e., not via an UTP. Preciously, $\frac{m}{2}$ bits will be used for communications from Alice to Bob, and the other half from Bob to Alice. In this case, Alice and Bob use two different bits for communications. And hence, the communications become sequential not simultaneous. Whereas, in our proposed scheme exploiting the same bit, Alice and Bob can communicate simultaneously. 

Moreover, for the first case, as we have to exploit $P\&M$ (preparation and measure) method, we can not guarantee no side channel leakage from the measurement devices of the legitimate parties. On the other hand, if we consider the proposed MDI scheme, it automatically certifies the robustness against any side channel attacks. Though there is no bitwise advantage between these two protocols (one, where Alice (res. Bob) uses $\frac{m}{2}$ bits for communications and other, the proposed protocol), but the current protocol offers better security.

For the first phase of the protocol, we consider the security analysis for the finite key length BB84 QKD presented by Tommamichel et al.~\cite{Tomamichel}. Resting on this we prove the entire security of our composed protocol.
We formalize our proposed protocol in {\bf{Algorithm $2$}}.

 The BB84 security analysis has been performed assuming the sources that emit weak coherent (light) pulses (WCP). However, the analysis can be extended for photon tagging~\cite{Norbert} and decoy states~\cite{Ma}.
 
 For the first phase of the protocol, one may perform decoy state BB84 to establish the secret key $b$ between Alice and Bob. However, for the second phase of the protocol i.e., for the actual quantum dialogue part one need not to use decoy states for the following reason.
 
Decoy states are exploited to resist Photon Number Splitting (PNS) attack~\cite{scarani04,Wang,norbert}. In a QKD protocol, after the exchange of qubits, the legitimate parties announce their measurement bases publicly. This public announcement is exploited in PNS attack.  
 
 Contrary to this, here, in MDI-QD phase the measurement bases have been pre-determined through BB84 QKD. Hence, there is no need to announce the bases publicly. Conditional on the event that the key distribution part has been performed securely, Eve can not perform PNS attack even if she gets several copies of a photon.
  
 The set-up for QD part is shown in the {\bf {Fig. $1$}}. 
 
\begin{figure*}[htbp]
{\centering
\includegraphics[width=.65\textwidth,trim = 1cm 6cm 1cm 1cm,clip]{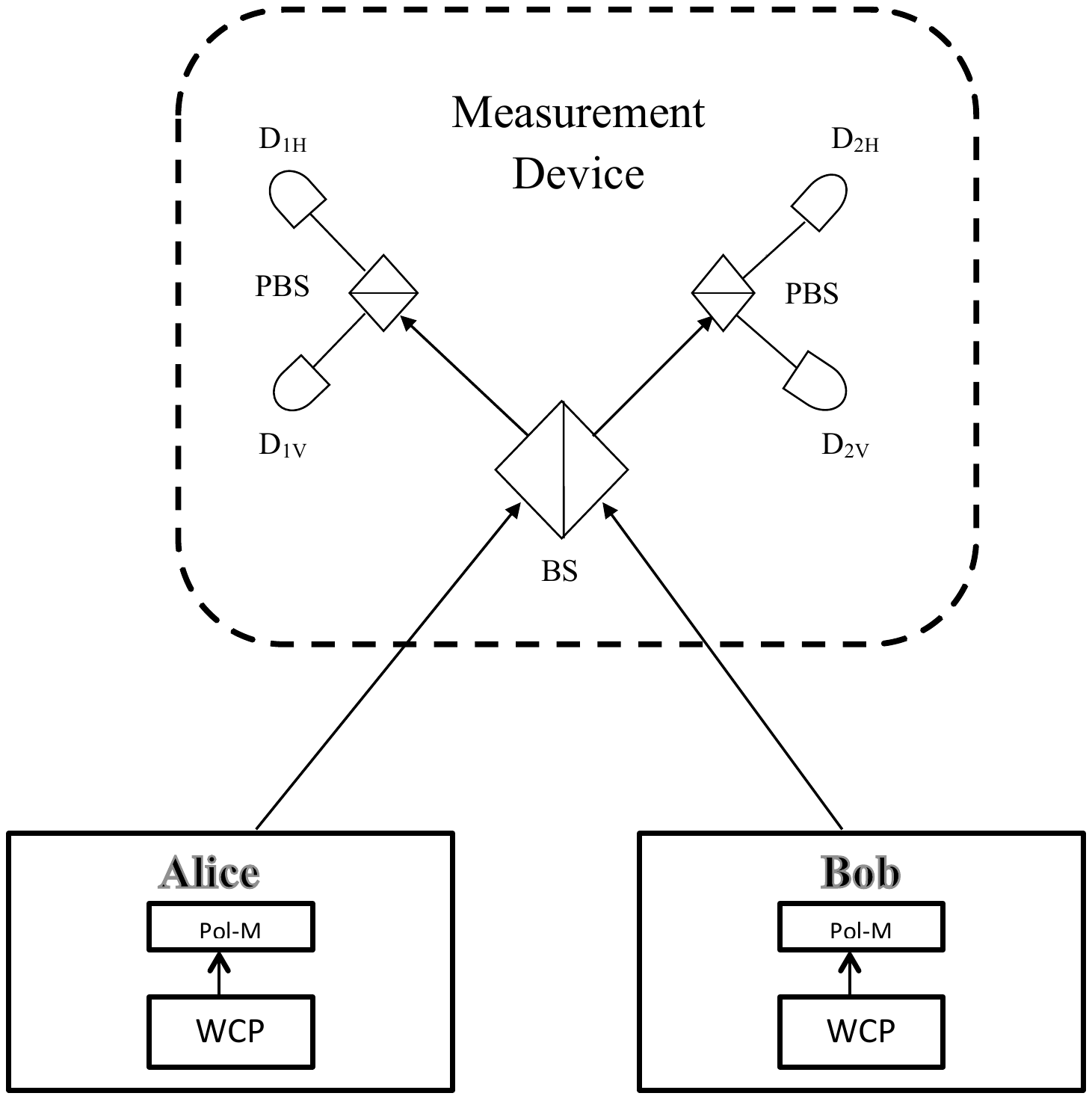}
\caption{The schematic diagram of MDI-QD phase is presented. Alice and Bob sends weak coherent pulses (WCP) in pre-determined polarizations. Those signals interfere at the 50 : 50 beam splitter (BS) which has a polarizing beam splitter (PBS) at each end.  Each PBS projects the incoming photons into either horizontal (H) or vertical (V) polarization states. A click in the single-photon detectors $D_{1H}$ and $D_{2V}$ or in $D_{1V}$ and $D_{2H}$ indicates a projection into the Bell state $\ket{\psi^{-}}  = \frac{1}{\sqrt{2}}(\ket{HV} -\ket{V H}$), while a click in $D_{1H}$ and $D_{1V}$ or in $D_{2H}$ and $D_{2V}$ implies a projection into the Bell state $\ket{\psi^{+}} = \frac{1}{\sqrt{2}}(\ket{HV}+\ket{VH})$.}}
\label{fig1}
\end{figure*}

This optical set-up is similar to the optical set-up considered in~\cite{Lo}. Here, we omit the intensity modulators (Decoy-IMs) which are used to prepare decoy states. It is commented in~\cite{Lo} that this optical set-up can only distinguish $\ket{\psi^{+}}$ and $\ket{\psi^{-}}$. However, this will not create any problem in our case. We will consider the cases where Eve declares $\ket{\psi^{+}}$ and discard the cases when Eve announces $\ket{\psi^{-}}$. The security analysis remains the same.

We should mention that the practical implementation of the proposed protocol is feasible exploiting current quantum technology. BB84 quantum key distribution devices are available in the international market~\cite{idq}. Experimental demonstration of MDI-QKD has also been reported~\cite{Lo1}. As the current protocol is the combination of these two, the experimental demonstration is possible in practice. Moreover, the protocol under consideration does not require any quantum memory which is still hard to achieve.


\begin{algorithm}[htbp]
\label{algo_yang}
\begin{enumerate}
\item Alice and Bob perform BB84 protocol to establish a $m$ bit key stream ($b$) between themselves.
\item Alice chooses a random bit string $a$ of length $m$ bits and Bob chooses another random bit string $a'$ of length $m$ bits.
\item Alice (Bob) prepares the qubits at her (his) end according to the following strategy.
\begin{itemize}
\item if $a_j (a'_j)=0$ and $b_j=0$, prepares $\ket{0}$.
\item  if $a_j (a'_j)=1$ and $b_j=0$, prepares $\ket{1}$.
\item  if $a_j (a'_j)=0$ and $b_j=1$, prepares $\ket{+}$.
\item  if $a_j (a'_j)=1$ and $b_j=1$, prepares $\ket{-}$.
\end{itemize} 
\item Alice (Bob) sends the prepared qubit to the third party Eve who in turn, measures the two qubits (one from Alice and one from Bob) in Bell basis and announces the result.
\item If Alice prepares $\ket{0}$ ($\ket{1}$) and the measurement result is either $\ket{\phi^{+}}$ or $\ket{\phi^{-}}$, Alice concludes with probability $1$ that the communicated bit of Bob is $0$ ($1$).
\item  If Alice prepares $\ket{0}$ ($\ket{1}$) and the measurement result is either $\ket{\psi^{+}}$ or $\ket{\psi^{-}}$, Alice concludes with probability $1$ that the communicated bit of Bob is $1$ ($0$).
\item If Alice prepares $\ket{+}$ ($\ket{-}$) and the measurement result is either $\ket{\phi^{+}}$ or $\ket{\psi^{+}}$, Alice concludes with probability $1$ that the communicated bit of Bob is $0$ ($1$).
\item If Alice prepares $\ket{+}$ ($\ket{-}$) and the measurement result is either $\ket{\phi^{-}}$ or $\ket{\psi^{-}}$, Alice concludes with probability $1$ that the communicated bit of Bob is $1$ ($0$).
\item Similar thing happens for Bob too.
\item Then they consider only the cases where the measurement results were $\ket{\phi^{-}}$ or $\ket{\psi^{+}}$ and discard the cases for $\ket{\phi^{+}}$ and $\ket{\psi^{-}}$.
\item To estimate the error introduced in this phase, Alice and Bob choose $\frac{\gamma m}{2}$ number of rounds, where $\gamma$ is a small fraction and reveal their respective guesses for these rounds. 
\item If the observed error lies below the predefined threshold value, they continue the protocol for the remaining $\frac{(1-\gamma) m}{2}$ rounds, otherwise they abort. 
 \end{enumerate}
 \caption{Protocol for MDI QSDC}
\end{algorithm}

\section{Security Analysis of the Proposed Protocol}
Before explaining the security analysis, we like to define some security notions required for the security proof of the suggested protocol. As the proposed protocol is based on BB84 QKD, we first define some security notions used in the security proof of BB84~\cite{Tomamichel}. Then we extend those security notions to QSDC. To the best of our knowledge, this is the first effort to formalise the security issues in this way in the domain of quantum secure direct communication.

In this regard, it should be mentioned that all the existing QSDC~\cite{qd0,qd1,qd2,qd3,qd4,qd5,qd6,qd7,qd8} and QD protocols~\cite{qd9,qd10,qd11,qd12,banerjee,shukla,pathak}) analyze the security issues showing the robustness against some specific type of attacks. For example, the security of a protocol is shown against Intercept and Resend Attack, Disturbance Attack or Modification Attack, Entanglement Measure Attack, Man-in the Middle Attack etc. In the present draft, instead of studying individual attack model we consider an approach which is exploited to prove the composable security of a protocol against any arbitrary attack. In this direction, we consider the entropic uncertainty relations presented in~\cite{Tomamichel3}.

It should be noted that protocols for direct quantum communication generally
suffer from the fact that error correction and privacy amplification
become problematic. However, error estimation phase is still there. We, here, show how the security of the proposed protocol is based on this error estimation phase. The information leakage to the adversary depends on the amount of quantum bit error rate (QBER) allowed by the protocol.

\subsection {\bf {Security Notions Used in QKD}} In this subsection we accumulate the security notions used to prove the composable security of QKD protocols. 
\begin{definition}{Correctness:}
A QKD protocol is said to be correct if for any strategy of an adversary, the key generated at Alice's end ($K_A$) is same with the key generated at Bob's end (${K_B}$). More preciously a QKD protocol is correct if and only if
$$K_A=K_B$$ 
\end{definition}

\begin{definition} {$\epsilon_{cor}$-correct:}
A QKD protocol is said to be $\epsilon_{cor}$-correct if it is $\epsilon_{cor}^{QKD}$ indistinguishable from a correct protocol, i.e., a protocol is $\epsilon_{cor}$-correct if 
$$\Pr [K_A \neq K_B] \leq \epsilon_{cor}^{QKD}$$ 
\end{definition}
\begin{definition}{$\Delta$-secret:} A key, generated from a QKD protocol, is called $\Delta$-secret from an adversary ($E$), if it is $\Delta^{QKD}_{sec}$ close to a uniformly distributed key that is uncorrelated with the adversary, i.e., if 
$$\min_{\rho_{E}}\frac{1}{2}||\rho_{K_{A}E}-\omega_{K_A}\otimes\rho_{E}||\leq \Delta^{QKD}_{sec},$$ 
where, $\rho_{{K_A}E}$ describes the correlation between the classical key ($K_A$)  of Alice and the eavesdropper ($E$), $\omega_{K_A}$ denotes the fully mixed state on $K_A$ and $\rho_{E}$ is the marginal state on the system of the adversary $E$.
\end{definition}
\begin{definition}{Secrecy:} A QKD protocol is said to be completely secret if for any attack strategy of an eavesdropper, $\Delta^{QKD}_{sec}=0$ whenever the protocol outputs a key. The key is called $\epsilon_{sec}$ if it is $\epsilon_{sec}^{QKD}$-indistinguishable from a secret protocol. Precisely, a protocol is $\epsilon_{sec}$-secret if it outputs $\Delta^{QKD}_{sec}$-secure keys with $(1-p_{abort})\Delta^{QKD}_{sec}\leq \epsilon_{sec}^{QKD}$, where $p_{abort}$ is the probability that the protocol aborts.
\end{definition}
\begin{definition}{Security:} A QKD protocol is called secure if it is both correct and secret. A protocol is $\epsilon$-secure if it is both $\epsilon_{cor}$-correct and $\epsilon_{sec}$-secret. In other words, a protocol is $\epsilon$-secure if
$$\epsilon_{cor}^{QKD}+\epsilon_{sec}^{QKD}\leq \epsilon^{QKD}.$$
\end{definition}

\subsection {\bf {Security Notions Defined for QSDC}} In this subsection we extend the above security notions for QSDC protocols. 
\begin{definition}{Correctness:}
The QSDC protocol is said to be correct if for any strategy of an adversary, Alice's (Bob's) guess $G_{A}$ ($G_{B}$) is same as Bob's (Alice's) communicated bits $C_B$ ($C_A$), i.e.,
$$\Pr(G_{A}=C_B)=\Pr(G_{B}=C_A)=1$$ 
\end{definition}
\begin{definition} {$\epsilon_{cor}$-correct:}
A QSDC protocol is said to be $\epsilon_{cor}$-correct if it is $\epsilon_{cor}^{QSDC}$ indistinguishable from a correct protocol, i.e., a protocol is $\epsilon_{cor}$-correct if 
$$\Pr [G_A \neq C_B]=\Pr[G_B \neq C_A] \leq \epsilon_{cor}^{QSDC}$$ 
\end{definition}
\begin{definition}{$\Delta$-secret:} A QSDC protocol is called $\Delta$-secret from an adversary ($E$), if the information extracted by the adversary from the protocol about the communicated bits of a legitimate party is $\Delta^{QSDC}_{sec}$ close to the random guess of the adversary, i.e.,

$$\min_{\rho'_{E}}\frac{1}{2}||\rho'_{{C_A}E}-\omega'_{C_A}\otimes\rho'_{E}||\leq \Delta^{QSDC}_{sec}$$
where, $\rho'_{{C_A}E}$ describes the correlation between the communicated bits of Alice and Eve, $\omega'_{C_A}$ is the completely mixed state on $C_A$ and $\rho'_{E}$ is the marginal state on the system $E$.

\end{definition}
\begin{definition}{Secrecy:} A QSDC protocol is said to be completely secret if for any attack strategy of an eavesdropper, $\Delta^{QSDC}_{sec}=0$. The protocol is called $\epsilon_{sec}$ if it is $\epsilon_{sec}^{QSDC}$-indistinguishable from a secret protocol. Precisely, a protocol is $\epsilon_{sec}$-secret if it outputs $\Delta^{QSDC}_{sec}$-secure communications with $(1-p_{abort}^{QSDC})\Delta^{QSDC}_{sec}\leq \epsilon_{sec}^{QSDC}$, where $p_{abort}^{QSDC}$ is the probability that the protocol aborts.

\end{definition}
\begin{definition}{Security:} A QSDC protocol is called secure if it is both correct and secret. A protocol is $\epsilon$-secure if it is both $\epsilon_{cor}$-correct and $\epsilon_{sec}$-secret. In other words, a protocol is $\epsilon$-secure if
$$\epsilon_{cor}^{QSDC}+\epsilon_{sec}^{QSDC} \leq \epsilon^{QSDC}.$$
\end{definition}

Let us now discuss the security related issues of our protocol with respect to the above definitions.

The security of our protocol completely depends on the security of $b$ which is generated exploiting BB84 QKD. Hence, our foremost job is to discuss the security of BB84 protocol which is the fundamental building block of our scheme. As we require finite length of the key, we consider the security proof of Tomamichel et al.~\cite{Tomamichel} where the security of BB84 for finite length key has been proven considering weak coherent light pulses. They claimed that the protocol is $\epsilon_{sec}$-secret if the secret key of length $l$ satisfies the following condition.

{\small
\begin{eqnarray*}
l \leq \lfloor n(q-h(Q+\mu)
-Leak_{EC}-\log_2(2/\epsilon_{cor})- 2\log_{2}(1/2 \bar{\epsilon})\rfloor
\end{eqnarray*}}
where, $n$ is the number of raw key bits, $q$ is the quality of the source which emits qubits, $Q$ is the tolerable quantum bit error rate (QBER), and $\mu$ is the
statistical deviation. $\mu$ is equal to $\sqrt{\frac{n+k}{nk}\frac{k+1}{k}\ln(1/\epsilon_{Q})}$, where $k$ is the number of bits chosen for error estimation and $\epsilon_{Q}$ is a negligibly small quantity in order of approximately $10^{-10}$. $Leak_{EC}+ 2\log(2/\epsilon_{cor})+2\log_{2}(1/2 \bar{\epsilon})$ is the maximal bits of information that are revealed to the eavesdropper during the protocol. $h(x)$ is the truncated binary entropy function, i.e., $h(x)=-(1-x)\log(1-x)-x\log(x)$ if $x\leq \frac{1}{2}$ but $1$ otherwise~\cite{supptomamichel}. 

 Now we prove the security of the second phase of our protocol, i.e., the actual quantum dialogue part. 
 The security of the entire protocol is solely based on how secure the key ($b$) is. For example, suppose the adversary succeeds to extract a single bit of the key ($b$). In this case, she will successfully extract the secret communications between the legitimate parties corresponding to that bit. Now, we assume that in the first phase of the protocol, Alice and Bob were able to extract a $\epsilon_{sec}$-secure key of length $m$. Conditional on this event, we now analyze the security of the quantum dialogue phase.

 Let the communicated bit by Alice be $C_A$. After the classical announcement of Bell basis measurement, let Bob's guess about $C_A$ be $G_B$ and the corresponding guess by Eve be $G_E$. 

Now, from the tripartite uncertainty relation~\cite{Tomamichel3}, we get that if $G_B$ is strongly correlated with $C_A$ in $\{\ket{0},\ket{1}\}$
basis or in $Z$ basis, then $G_E$ should be completely uncorrelated with $C_A$ in $\{\ket{+},\ket{-}\}$ basis or in $X$ basis. 
Hence, we can write
\begin{eqnarray*}
 H(X|G_E)+ H(Z|G_B) = -\log_{2}{c}
 \end{eqnarray*}
 where, $c$ is the maximum overlap of the two bases, i.e., $c=\underset{u,v}{\max}|\langle{\theta^{u}_{X}}|{\theta^{v}_{Z}}\rangle|^2$, where $\ket{\theta_{X}^{u}}$ and $\ket{\theta_{Z}^{v}}$ are the eigenvectors of two measurement operators $X$ and $Z$ respectively. Here, $-\log_2{c}=1$.
 
 As Alice and Bob always choose the same basis which has been pre-determined through BB84 QKD, they should be completely correlated (or anti-correlated) in each run of the protocol. Conditional on the event that the key distribution part is $\epsilon_{sec}$-secure and from the principle of no-cloning~\cite{wootters,dieks}, we get that the information gain by Eve is only possible by introducing some error in the channel. 
 
  Let the Quantum Bit Error Rate (QBER) at $\{\ket{0},\ket{1}\}$ basis be $Q'$. Hence, we can write,  $H(Z|G_B) = H(Q')$ \footnote{By the abuse of notation, we can write $H(U)=H(p)$~\cite{cover}, where, $U$ is a discrete random variable with corresponding probability mass function $p$.}.
 
 As the protocol is symmetric i.e., $H(X|G_B)=H(Q')$, the total uncertainty of Bob about $C_A$ can be written as 
 \begin{eqnarray*}
 H(C_A|G_B)=2H(Q')
 \end{eqnarray*}
 
 Now, if Eve was completely uncorrelated with $C_A$, then $H(C_A|G_E)= H(C_A)=1$. However, we assume that mounting an arbitrary attack Eve will be successful to extract some information about $C_A$. In exchange she has to introduce some error in the channel. This disturbance is quantified by the uncertainty of Bob about $C_A$. And hence, the total uncertainty of Eve about $C_A$ becomes 
 \begin{eqnarray*}
 H(C_A|G_E) &=& H(C_A)-H(C_A|G_B)\\
 &=& 1-2H(Q')
 \end{eqnarray*}
 Now, from the Serfling lemma~\cite{Serfling} and its corresponding corollary~\cite{Lim} we can immediately conclude that for the remaining set of $\frac{(1-\gamma) m}{2}$ bits the QBER in $\{\ket{0}, \ket{1}\}$ basis must be in the range $[Q'-\nu, Q'+\nu]$, where $\nu$ is the statistical deviation and $\nu=\sqrt{\frac{(\gamma m+2)}{\gamma^2(1-\gamma)m^2}\ln{\frac{1}{\epsilon^{QSDC}}}}$. Thus, the maximum mutual information between Alice and Eve becomes
 \begin{eqnarray*}
 I(C_A; G_E)&=& H(C_A)- H(C_A|G_E)\\
 &\leq & 2 H(Q'+\nu).
 \end{eqnarray*}
 
 Whether the parties abort the protocol completely depends on the error estimation phase. If QBER ($Q'$) is less than the tolerable bit error rate suggested by the protocol, then $p_{abort}^{QSDC}=0$. That means conditional on the event that the error estimation phase passed, we can write
  $$\Delta^{QSDC}_{sec} \leq \epsilon^{QSDC}_{sec}.$$ 
  That means the protocol is $\epsilon_{sec}$-secure as 
$$\min_{\rho'_{E}}\frac{1}{2}||\rho'_{{C_A}E}-\omega'_{C_A}\otimes\rho'_{E}||\leq \Delta^{QSDC}_{sec}\leq \epsilon^{QSDC}_{sec}.$$

Now, we analyze the correctness of our protocol. If the announced result was correct then Alice as well as Bob can predict the communicated bits by her or his counter parts with probability $1$ (see Table $1$). Precisely, 
$$\Pr(G_A=C_B)=1.$$
However, due to the noise of the channel $\Pr(G_A = C_B)$ should not be $1$. 
Rather, we can write 
\begin{eqnarray*}
\Pr(G_A\neq C_B)& \leq & (Q'+\nu)
 \end{eqnarray*}
 To ensure the $\epsilon_{cor}$-correctness of the protocol, we have to restrict the value of $Q'$ by $\epsilon_{cor}^{QSDC}$, i.e., $ (Q'+\nu) \leq  \epsilon_{cor}^{QSDC}$. 
Now, we choose $\epsilon^{QSDC}_{cor} \approx \epsilon^{QSDC}_{sec} \approx \frac{\epsilon^{QSDC}}{2}$. Finally, if we assume that $\epsilon^{QSDC}$ is proportional to $\epsilon^{QKD}$, then we can say that our protocol is $\epsilon$-secure as the following holds. 
\begin{eqnarray*}
\epsilon^{QSDC}_{sec}+\epsilon^{QSDC}_{cor} & \approx &\epsilon^{QSDC} \approx \epsilon^{QKD}
 \end{eqnarray*}

 However, one may find that it is not sufficient for the security as there is no privacy amplification process like QKD. Thus, if the tolerable bit error rate is too large, the information of the communicated bits in the dialogue phase will be leaked too much. On the other hand, if the tolerable bit error rate is set to be zero, then the protocol will become impractical as it requires infinite number of qubits. To remove such ambiguity, in the following section, we give a detailed study of information leakage in   existing QD protocols and compare that with the information leakage of the proposed protocol.

 \section{Comparison with Existing QD Protocols}\label{com} We start with the very first protocol by Nguyen~\cite{qd9}. In this protocol Bob starts with $N$ entangled pairs initially prepared in state $\ket{\psi^{+}}$. Bob retains one half of each entangled pair and sends other half to Alice. There are two modes of operations. One is called control mode (CM) and another is called message mode (MM). Alice can randomly shifts between these two modes. CM mode is performed to check the error in the channel. If the error rate is lower than the pre-defined threshold value they will continue the protocol. Otherwise, they will abort.

In MM mode Alice chooses any one of the operators from the set $\{I, \sigma_x, i\sigma_y, \sigma_z\}$. Each operation is encoded by two classical bits. For example, $I$ is encoded by $00$, $\sigma_x$ is encoded by $01$, $i \sigma_y$ is encoded by $10$ and $\sigma_z$ is encoded by $11$. Alice encodes her qubit by applying the operator on it and returns that back to Bob. Bob then encodes the same qubit by choosing one of the above operators. After that he measures the encoded qubit and the qubit retained by him in Bell basis and announces the result publicly. Depending on the announcement Bob and Alice decode the messages of their respective counter-parts.

Now, suppose after Bell measurement Bob obtains $\ket{\psi^{-}}=\frac{1}{\sqrt{2}}(\ket{01}-\ket{10})$. The announcement of this measurement result provides the following informations.
\begin{itemize}
\item Bob applies $I$, Alice applies $\sigma_z$ or vice versa
\item Bob applies $\sigma_x$, Alice applies $i \sigma_y$ or vice versa
\end{itemize}
Thus from the announcement, the adversary (Eve), immediately comes to know that the communicated bits between Bob and Alice must be one of the followings. 
\begin{itemize}
\item $(00,11), (01,10), (10, 01), (11, 00)$.
\end{itemize}
 However, from the perspective of cryptography, the protocol would be completely secure if the situation before the announcement and after the announcement remains indistinguishable. Precisely, after the announcement the communicated bits between Bob and Alice should still be a random guess to Eve. That is Eve should only know the communicated bits can be any of the following exhaustive possibilities.
 \begin{itemize}
\item $(00, 00), (00, 01), (00,10), (00,11),
 (01, 00), (01, 01),
 (01, 10), (01,11), (10, 00),\\
 (10, 01),
 (10, 10), (10,11), 
(11, 00), (11, 01), (11, 10), (11, 11).
$ 
\end{itemize}
     
Hence the incorrect usage of classical communication provides Eve $-4\times \frac{1}{4} \log{\frac{1}{4}}= 2$ bits of information among $4$ bits without introducing any noise in the channel. Unfortunately, all the variants~\cite{qd10,qd11,qd12,banerjee,shukla,pathak} of~\cite{qd9} suffer from such information leakage.

On the other hand, we design our protocol in such a way so that we can remove this inherent information leakage through the protocols. Close observation of table $1$ revels that if the measurement result is either $\ket{\phi^{+}}$ or $\ket{\psi^{-}}$, Eve immediately comes to know that the communicated bits between Bob and Alice will be either $\{00, 11\}$ or $\{01, 10\}$ respectively. So in these two cases $-2\times\frac{1}{2}\log{\frac{1}{2}}=1$ bit of information is gained by Eve without introducing any noise in the channel. However, if she measures $\ket{\phi^{-}}$ or $\ket{\psi^{+}}$, she does not get any extra information regarding the communicated bits as each of the $4$ strings of two bits are equally probable now. This is why we discard the cases where Eve measures $\ket{\phi^{+}}$ or $\ket{\psi^{-}}$ to prevent inherent information leakage.

The existing QD protocols allow this inherent information leakage which is equal to the half of the bits communicated. Explicitly, if Alice and Bob exchange $4$ bits of classical information, then the protocols allow $2$ bits of information to Eve. If Alice and Bob exchange $2$ bits of classical information, then the protocols allow Eve to gain $1$ bit of information without any effort. Moreover, setting some positive threshold value for QBER the protocols provide additional information to Eve. As there is no privacy amplification phase, there is no way to discard this information leakage to the adversary. 

Contrary to this, in the present protocol there is no inherent information leakage. To gain any information regarding the communication Eve has to introduce some noise in the channel. Considering the same threshold value of QBER as the existing protocols, our scheme leaks less information and hence offers better security.

\section {Usage of Symmetric Cipher} In this section we explore the advantage for exploiting symmetric cipher in the current QD protocol. 
Suppose, we want to continue the dialogue for a very long time, say for $T$ times. However, we require $M$-bits security against quantum adversary while $T>>2M$. In that case, it is wiser to use the construction of a symmetric-cipher to reduce the `cost' of the protocol. Here, `cost' means the number of qubits used and the number of communications (classical as well as quantum) needed for this protocol. 

To generate a $T$-bits key-stream using a symmetric cipher, we can exploit a bitstream of smaller length (say, $2M$ bits whose security is reduced to $M$ bits against quantum adversary~\cite{grover}). This smaller bit string is known as `seed'. The secrecy of the cipher depends on the secrecy of the seed used. In the current protocol, this initial seed is generated between the legitimate parties exploiting BB84 protocol and then both the parties may use a secure stream cipher~\cite{SoK} to generate 
longer secret key-stream at both ends.

For example, consider that we need $128$ bits security, but the quantum dialogue may be repeated for $512$ times. In this case if we do not use symmetric cipher, we have to generate $2048$ bits long steam through BB84. This is because, half of the bits which correspond to the measurement results $\ket{\phi^{+}}$ and $\ket{\psi^{-}}$ will be discarded and approximately half of the remaining bits will be taken for error estimation. So to initiate the key distribution part we require approximately $8192$ number of qubits ($4$ times of the key bits). However, if we use symmetric cipher considering the seed of $256$ bits ($2$ times of $128$-bits), we need approximately $1024$ qubits and hence that much quantum communications. In practice, to generate a finite length key we require far more qubits~\cite{Tomamichel} estimated above.

In this regard, we can recall Table $1$ of~\cite{Tomamichel}. Assuming QBER$=1\%$ and security rate $\epsilon/l=10^{-14}$ we get the key generation rate $(r)=11.7\%$. From this we can conclude that to generate $2048$-bit $\epsilon$ -secure key ($b$) approximately $17,504$ raw key bits are required. That means we have to start with four times of $17,504$ qubits which is equal to $70,016$ qubits and that amount of quantum communications. 
On the other hand, to generate $256$ bit seed we require approximately $2188$ raw key bits. In this case, we start with $8752$ qubits and that amount of quantum communications. Hence, in term of reduction of the cost it will be wiser to choose a symmetric cipher to generate the bit stream $b$.

The existing QKD schemes have the unique property that those are universally composable (UC)~\cite{canneti}. This universally composable security of QKD protocols are well proven~\cite{renner}. This implies that QKD can be combined with any other UC resulting a composed protocol which is also a UC~\cite{TCS}. That is why, in our current protocol, we can compose BB84 QKD with a symmetric cipher~\cite{SoK} to generate $b$.

\section{Discussion and Conclusion}
\label{con}
In this current draft, we propose a measurement device independent quantum dialogue protocol. Measurement device independent quantum key distribution~\cite{Lo} has drawn a huge attention as it can resist the side-channel attacks under certain assumptions. Motivated by the idea of~\cite{Lo}, we propose a similar scheme for quantum dialogue. 

One may wonder why we would not replace the key distribution part of the protocol by MDI-QKD to provide fully measurement device independence security. In this regard, we like to mention that BB84 QKD has been proven to be one-sided device independent~\cite{Tomamichel13}. In the security analysis in~\cite{Tomamichel13} we have to put trust only on the source which emits the entangled pairs. Measurement devices have been treated as a black box. However, one may replace the BB84 part by MDI-QKD. The security proof remains same for both the cases.

Very recently, the experimental demonstration of QSDC protocol with quantum memory has been reported~\cite{zhang}. Being a subset of QSDC, analysing the security issues for QD protocols becomes very important.

Our QD scheme does not require quantum memory. As soon as the third party (TP) gets the qubits (one from Alice and one from Bob), she measures those in the Bell basis and announces the measurement results. Depending on the announcement Alice and Bob come to know the communicated bits by their respective counter-parts. Hence, the current protocol is inherently robust against any type of quantum memory attacks which might be carried on the QSDC protocols which require quantum memory. 

All the existing QD protocols suffer from inherent information leakage via classical communications about the measurement outcomes~\cite{qd13,qd14}. We design our protocol in such a way so that it becomes robust against such information leakage. So, for practical implementation, our scheme offers better security over the existing QSDC and QD protocols.



\begin{thebibliography}{}
\bibitem{Shor}
   P. W. Shor,
    {\it Foundations of Computer Science (FOCS) 1994}, 124--134, IEEE Computer Society Press, 1994
   \bibitem{Superden}
    C. Bennett,  S. Wiesner, 
    {\it Phys. Rev. Lett.}, {\bf 69 (20)}, 2881, 1992
    \bibitem{qd0}
    G. L. Long, X. S. Liu, 
    {\it Phys. Rev. A}, {\bf 65}, 032302, 2002
    \bibitem{qd1}
    K. Bostr\"{o}m, T. Felbinger,
    {\it Phys. Rev.
Lett.}, {\bf 89}, 187902, 2002
    \bibitem{qd2}
    F. G. Deng, G. L. Long, X. S. Liu,
    {\it Phys. Rev. A}, {\bf 68}, 042317, 2003
    \bibitem{qd3}
    F. G. Deng, G. L. Long, 
    {\it Phys. Rev. A},
{\bf 69}, 052319, 2004
    \bibitem{qd4}
    M. Lucamarini, S. Mancini,
    {\it Phys. Rev.
Lett.}, {\bf 94}, 140501, 2005
    \bibitem{qd5}
   C. Wang, F. G. Deng, Y. S. Li  et al.,
   {\it Phys. Rev. A}, {\bf 71}, 044305, 2005
   
   \bibitem{qd6}
   C. Wang, F. G. Deng, G. L. Long, 
  {\it Opt. Commun.}, {\bf 253}, 15--20, 2005

\bibitem{qd7}
X. H. Li, F. G. Deng, H. Y. Zhou, 
{\it Phys. Rev. A}, {\bf 74}, 054302, 2006
\bibitem{qd8}
X. H. Li, C. Y. Li, F. G. Deng et al.,
{\it Chin. Phys.}, {\bf 16}, 2149--2153, 2007
\bibitem{qd9}
B. A. Nguyen,
{\it Phys. Lett. A}, {\bf 328}, 6--10, 2004
\bibitem{qd10}
Z. X. Man, Z. J. Zhang, Y. Li,
{\it Chin. Phys. Lett.}, {\bf 22}, 22--24, 2005
\bibitem{qd11}
 X. Ji, S. Zhang, 
 {\it Chin. Phys.}, {\bf 15}, 1418--1420, 2006
 \bibitem{qd12}
 Z. X. Man, Y. J. Xia, B. A. Nguyen,
 {\it J Phys. B: At. Mol. Opt. Phys}, {\bf 39}, 3855--3863, 2006
  
  \bibitem{banerjee}
 A. Banerjee, C. Shukla, K. Thapliyal, A. Pathak, P. K. Panigrahi, 
 {\it Quant. Inf. Process.} {\bf 16}, 49, 2017
 
  \bibitem{shukla}
  C. Shukla, V. Kothari, A. Banerjee,  A. Pathak, 
  {\it Phys. Lett. A}, {\bf 377}, 518, 2013

\bibitem{pathak}
C. Shukla, K. Thapliyal, A. Pathak, 
arXiv:1702.07861

  
  
\bibitem{zhang}
W. Zhang, D. S. Ding, Y. B. Sheng, L. Zhou, B. S. Shi, G. C. Guo,
{\it Phys. Rev. Lett.}, {\bf 118}, 220501,  2017

\bibitem{qd13}
Y -G. Tan, Q -Y. Cai,
{\it Int. J. Quant. Inf.}, {\bf 6 (2)}, 325, 2008
\bibitem{qd14}
F. Gao, F -Z. Guo, Q -Y. Wen, F -C. Zhu, 
Science in China Series G: Physics, Mechanics \& Astronomy, {\bf 51 (5)}, 559--566, 2008



 \bibitem{Lo}
H. K. Lo, M. Curty, B. Qi,
{\it Phys. Rev. Lett.}, {\bf 108}, 130503, 2012

\bibitem{bb84}
 C. H. Bennett, G. Brassard, 
In Proceedings o f IEEE International Conference on Computers, Systems and Signal Processing, {\bf 175}, 8, 1984

\bibitem{Tomamichel}
M. Tomamichel, C. C. W. Lim, N. Gisin, R. Renner,
{\it Nat. Commun.} {\bf 3}, 634, 2012

\bibitem{Norbert}
N. L{\"u}tkenhaus,
{\it Phys. Rev. A}, {\bf 61}, 052304, 2000

\bibitem{Ma}
H. -K. Lo, X. Ma, K. Chen,
{\it Phys. Rev. Lett.}, {\bf 94}, 230504, 2005


\bibitem{scarani04}
V. Scarani, A. Ac\'{i}n, G. Ribordy, N. Gisin,
{\it Phy. Rev. Lett.}, {\bf 92}, 057901, 2004
\bibitem{Wang}
X -B Wang,
{\it Phy. Rev. Lett.}, {\bf 94}, 230503, 2005
\bibitem{norbert}
N. L\"{u}tkenhaus, M. Jahma,
{\it New. J. Phys.}, {\bf 4}, 44-1, 2002

\bibitem{idq}
http://www.idquantique.com

\bibitem{Lo1}
Z. Tang,  Z. Liao, F. Xu, B. Qi, L. Qian, H-K Lo,
{\it Phy. Rev. Lett.}, {\bf 112}, 190503, 2014

\bibitem{Tomamichel3}
M. Tomamichel, E. H{\"a}nggi,
{\it J. Phys. A}, {\bf 46}, 055301, 2013.



\bibitem{supptomamichel}
M. Tomamichel, C. C. W. Lim, N. Gisin, R. Renner,
Supplementary Material,
\url {http://www.nature.com/ naturecommunications}, 2012

\bibitem{cover}
T. M. Cover, J. A. Thomas,
Elements of Information Theory, 
John Willey and Sons Inc., UK, Reprint 2009

\bibitem{wootters}
W. K. Wootters, W. H. Zurek,
{\it Nature}, {\bf 299}, 802803, 1982
 
 \bibitem{dieks}
 D. Dieks,
 {\it Phys. Lett.  A}, {\bf 92 (6)}, 271272, 1982
 
 \bibitem{Serfling}
R. J. Serfling,
{\it Ann. Stat.}, {\bf 2}, 39, 1974
\bibitem{Lim}
C. C. Wen Lim, C. Portmann, M. Tomamichel, R. Renner, N. Gisin,
{\it Phys. Rev. X}, {\bf 3}, 031006, 2013

 
\bibitem{grover}
 L. K. Grover, 
 Proceedings of the Twenty Eighth Annual ACM Symposium on Theory of Computing, 212--219, ACM, New York, NY, USA, 1996
 \bibitem{SoK}
S. Ruhault, 
IACR Transactions on Symmetric Cryptology, {\bf 2017 (1)}, 506--544, 2017
 
\bibitem{canneti}
R. Canetti, 
FOCS 2001, 136--145, 2001
\bibitem{renner}
 R. Renner, 
 {\it Int. J. Quantum
Inform.}, {\bf 6 (1)}, 127, 2008,
 eprint arXiv:quant-ph/0512258

\bibitem{TCS}

R. All\'{e}aume et al.,
{\it Theo. Com. Sci.}, {\bf 560}, 62--81, 2014

\bibitem{Tomamichel13}
M. Tomamichel, S. Fehr, J. Kaniewski, S. Wehner,
Eurocrypt 2013, {\bf 7881}, 609--625, 2013
\end{thebibliography}


\end{document}